\documentclass[preprint]{aastex}

\usepackage{graphicx}
\usepackage{epsfig,latexsym,amsmath}
\usepackage{natbib}
\usepackage{color}
\newcommand{\red}[1]{\colorbox{red}{#1}}
\newcommand{\yellow}[1]{\colorbox{yellow}{#1}}

\shorttitle{Timing Noise}
\shortauthors{Perrodin et al.}

\begin{document}

\title{Timing Noise Analysis of NANOGrav Pulsars}

\author{
D. Perrodin\footnote{INAF-Osservatorio Astronomico di Cagliari, Via della Scienza 5, 09047 Selargius (CA), Italy. Email: delphine@oa-cagliari.inaf.it} $^{,}$\footnote{Franklin \& Marshall College, 415 Harrisburg Pike, Lancaster PA 17604, USA} ,
F. Jenet\footnote{Center for Gravitational Wave Astronomy, University of Texas at Brownsville, Brownsville TX 78520, USA} , 
A. N. Lommen\footnotemark[2] , 
L. S. Finn\footnote{Department of Physics, The Pennsylvania State University, University Park PA 16802, USA} ,  
P.  B. Demorest\footnote{National Radio Astronomy Observatory, Charlottesville VA 22903, USA} ,
R. D. Ferdman\footnote{Department of Physics, McGill University, 3600 University Street, Montreal, Quebec H3A 2T8, Canada} , 
\\
M. E. Gonzalez\footnote{Department of Physics and Astronomy, University of British Columbia, Vancouver BC V6T 1Z1, Canada} , 
D. J. Nice\footnote{Department of Physics, Lafayette College, Easton PA 18042, USA} ,
S. Ransom\footnotemark[5] , 
I. H. Stairs\footnotemark[7]
}


\begin{abstract}

We analyze timing noise from five years of Arecibo and Green Bank observations of the seventeen millisecond pulsars of the North-American Nanohertz Observatory for Gravitational Waves (NANOGrav) pulsar timing array. The weighted autocovariance of the timing residuals was computed for each pulsar and compared against two possible models for the underlying noise process. The first model includes red noise and predicts the autocovariance to be a decaying exponential as a function of time lag. The second model is Gaussian white noise whose autocovariance would be a delta function. We also perform a ``nearest-neighbor" correlation analysis. We find that the exponential process does not accurately describe the data. Two pulsars, J1643-1224 and J1910+1256, exhibit weak red noise, but the rest are well described as white noise. The overall lack of evidence for red noise implies that sensitivity to a (red) gravitational wave background signal is limited by statistical rather than systematic uncertainty. In all pulsars, the ratio of non-white noise to white noise is low, so that we can increase the cadence or integration times of our observations and still expect the root-mean-square of timing residual averages to decrease by the square-root of observation time, which is key to improving the sensitivity of the pulsar timing array.  

\end{abstract}

\maketitle

\section{Introduction}

Millisecond pulsars (MSPs) are the most stable rotators in the universe:  in the best pulsars, we measure the arrival times of radio pulses with a precision of tens of nanoseconds. Over several years, the period of an MSP can be determined to one part in $10^{15}$. Due to their outstanding rotational stability, one could detect gravitational waves (GWs) from distant supermassive black hole binaries using an array of MSPs: a pulsar timing array (PTA) \citep{fosterbacker}. A large number of supermassive black hole binaries in the universe are believed to form a stochastic background of nanohertz GWs \citep{jaffebacker}. A GW passing between Earth and a pulsar would cause an offset in the arrival times of the radio pulses coming from the pulsar. This effect is correlated between pulsars located at different positions in the sky, allowing us to differentiate GWs from other phenomena \citep{hellingsdowns}. In order to detect this correlation pattern, we need an array of pulsars that are timed to a high precision. The ``timing precision" is quantified as the root-mean-square (RMS) of the timing residuals, where timing residuals are the differences between observed times-of-arrival of the pulses (TOAs) and the expected TOAs based on our best knowledge of the physics of pulsars and the propagation of radio waves through the interstellar medium. Non-random patterns in the residuals, in particular patterns with non-white power spectra, are referred to as ``timing noise''. \citet{jenet2005} predicted that in order to detect GWs, a PTA requires at least 20 MSPs to be timed with an RMS of 0.1 to 0.5 $\mu s$ over five to ten years. So far the required precision is being reached for a handful of pulsars \citep{demorest, rutger}. However with improvements to the observed systems across Australia, North America and Europe, one can expect a direct detection of GWs in the nanohertz frequency range using PTAs within the next decade \citep{demorest2009}.

While the main purpose of a pulsar timing array is to make a direct detection of GWs, one can immediately use GW detectors to place an upper limit on the amplitude of the stochastic background of GWs. Different detectors place different upper limits  on the characteristic strain. The best ever constraints on the strain of the stochastic background of GWs in the nanohertz regime were obtained by the European Pulsar Timing Array, whose analysis found an upper limit of: $h_c (1 \, \text{yr}^{-1}) < 6 \times 10^{-15}$ at the $95\%$ confidence level \citep{rutger}, while NANOGrav recently set an upper limit of $7 \times 10^{-15}$ \citep{demorest}. In order to improve the sensitivity of a PTA, we could: add more stable MSPs to the array (with a low level of timing noise); increase the total data span of our observations; increase the cadence of observations or the integration time on each pulsar (i.e., require more telescope time), or improve the way we analyze the data to increase the precision of the observations. However, increasing the cadence or the integration time only makes sense when the noise in the arrival times of the pulses is uncorrelated or ``white", i.e. when the power spectrum of timing residuals would show the power to be flat across all frequencies. In this case, we expect the RMS of timing residual averages to decrease as $1/\sqrt{N}$, where N is the total number of observations \citep{handzo}. However another noise term can appear in data observed with longer timespans (5-10 years). Prevalent at low frequencies, it is called ``red noise". When red noise is present, the timing precision will not increase at this rate, if at all. 

While MSPs are more stable than younger pulsars, a handful of them do exhibit red noise \citep{shannoncordes}. As we approach the timing precision required to achieve a direct detection of GWs, it may be that the phenomena seen in the younger pulsars are also apparent in more MSPs. GWs are expected to induce correlated red noise in the timing residuals of pairs of pulsars. Understanding the origin of red noise and subsequent mitigation would allow us to improve our pulsar timing sensitivity and allow us to differentiate GWs from timing noise. To mitigate red noise, one needs to accurately model its behavior. While still poorly understood, possible sources of red noise have been identified \citep{hobbs3}. It can be either intrinsic to the pulsar or due to the changing influence of the interstellar medium (ISM) on the propagation of the pulses. Intrinsic sources of red noise which have been identified or proposed include:  the superfluid interior of the neutron star affecting the rotation; variations in the magnetosphere affecting the spin-down torque, resulting in a random walk in pulse frequency or phase; pulse shape variation as a function of both time and observing frequency; unknown orbital companions (e.g., asteroids or planets) \citep{shannoncordes}.

How can we identify red noise? While the presence of strong red noise is obvious in timing residuals, identifying and quantifying a red noise component can prove difficult, especially in the case of weak red noise.  In this paper, we wrestle with quantitative ways to assess the presence of red noise in MSPs. Differentiating sources of noise can be done by computing a power spectrum of timing residuals. In the case of white noise, we expect a flat spectrum at all frequencies, but in the case of red noise, there is increasing power at lower frequency. However the challenge with pulsar data in particular, is that the data are non-uniformly sampled. In the case of NANOGrav data taken at the Arecibo and Green Bank telescopes \citep{demorest}, the data are taken over intervals of less than 30 minutes, once every month or so, with occasionally skipped months, and with varying numbers of days (between 20 and 40) between observations. 

The Fourier transform is only adequate for uniformly-sampled data. The problem of spectral leakage also makes Fourier transforms inappropriate for pulsar timing noise and GW background analysis \citep{coles}. It is possible to compute a Lomb-Scargle periodogram to compute a power spectrum \citep{jenet2004}. However the Lomb-Scargle method is not adequate for steep spectra \citep{coles}. Bayesian methods have been proposed to estimate red noise in pulsars \citep{rutger_yuri}. A bayesian analysis of timing noise in NANOGrav pulsars will be discussed in the work of J. Ellis et al. (2013, in preparation). \citet{coles} proposed to use the Cholesky method to compute a power spectrum of timing residuals and characterize the noise. The equivalent approach in the time domain consists in computing the autocovariance or autocorrelation function of timing residuals and, in the case of weak read noise, fitting an exponential. In this paper, we characterize the timing noise in 17 NANOGrav pulsars from \citet{demorest} by studying the autocovariance of timing residuals in the time domain. In Section 2.1 we compute the ``nearest-neighbor" autocovariance, similarly to the method used in the ``CheckWhite" plugin (made available by Hobbs et al. in the TEMPO2 distribution).  In Section 2.2 we compute the weighted autocovariance of timing residuals, and present an exponential fit  similar to the one described in \citet{coles} and implemented in the ``SpectralModel'' plugin (also available in the TEMPO2 distribution). We then describe a delta-function statistic and the calculation of a ``time factor'', which both help us determine the presence of non-white noise in the timing residuals.

\section{Methods for timing noise identification}

We used timing data from the 17 MSPs described in \citet{demorest}. Pulsars were monitored monthly for 5 years (from 2005 until 2010) at the 305-m NAIC Arecibo Observatory (AO) and the 100-m NRAO Green Bank telescope (GBT). The data consist of nearly 30 minutes of observations each month for each pulsar and frequency band. The frequency bands used at Arecibo were: 327, 430, 1400 and 2300 MHz. At the Green Bank telescope, they were: 820 and 1400 MHz. In this data set, the dispersion measure (DM) was fit at every epoch, which tends to lower the RMS of timing residuals found at the lower frequencies. Fitting for DM also tends to remove the longest-wavelength signals, due to covariance between DM variations and timing noise. Additionally, fitting pulsar parameters removes power from the autocovariance, especially at long time lags, due to removal of a quadratic from the residuals when fitting for period and period derivative. The uncertainties used are the original uncertainties from TOA calculations, and were not adjusted to give $\chi^2$ values of 1. Additional details about data acquisition and timing methods, as well as timing residuals for each pulsar are available in \citet{demorest}.

In the following methods (Sections 2.1 and 2.2), we used NANOGrav TOAs and ephemeres \citep{demorest} and ran TEMPO2 to obtain timing residuals. We worked with each major frequency band separately (i.e. 800, 1400 or 2300 MHz, etc.). Indeed, looking at autocovariances at different lags would be misleading if adjacent observations were taken for different frequency bands. We then daily-averaged the obtained residuals using weighted averages. \citet{demorest} data contain about 30 measurements (TOAs) each day in 4 MHz sub-bands, and computing correlations using multiple observations each day would falsely reveal correlations at short time scales, which are not the time scales that we are interested in.

In order to characterize timing noise in MSPs, we work in the time domain. In Section 2.1, we study a special case: the correlation of adjacent timing residuals, which we call ``nearest neighbor'' autocovariance $cov(\text{n.n})$. In Section 2.2, we compute the weighted autocovariance of timing residuals at all lags, and in Section 2.2.1, we study the exponential fit method to quantify the presence of red noise. In this method, the autocovariance is fit by a decaying exponential. In Section 2.2.2, we study a delta-function statistic to measure any deviations from white noise. Indeed in the case of white noise, we expect the autocovariance to behave like a delta function: $cov(\tau) = \sigma^2 \delta(\tau)$, where $\sigma^2$ is the variance of timing residuals. In the case of red noise, we expect a deviation from a delta function. Finally in Section 2.2.3, we present the computation of a ``time factor'' that helps us determine the level of non-white noise in the timing residuals.

\subsection{``Nearest neighbor'' autocovariance}

When computing the autocovariance of timing noise, we expect zero covariance at non-zero lags in the case of white noise. However in the case of non-white noise, some non-zero covariance should appear at the first non-zero lag. This is the method implemented in the ``CheckWhite" plugin. This method involves computing the ``nearest neighbor'' autocovariance. For data that are uniformly-sampled, this lag would be the same for all data, i.e. a lag of 1 in whichever unit the time sample is. However since we are dealing with non-uniformly sampled data, an interesting nuance is that here the autocovariance is computed using each closest pair of observations, where the time lag separating them will not be constant. This is the quantity:
\begin{equation}
cov(\text{n.n.})= \frac{1}{N-1} \sum_{i=1}^{N-1} {R_i \, R_{i+\text{n.n.}}},
\end{equation}
where $\text{R}_i$ and $\text{R}_{i+\text{n.n}}$ are adjacent (daily-averaged) timing residuals, and N is the number of residuals in the sample (note: here we calculated the  ``unweighted'' autocovariance).
Following equation (1), we computed, separately for each frequency band, the ``nearest neighbor" autocovariance $cov(\text{n.n})$, which measures the amount of correlation in the timing residuals. The probability of chance occurrence of the measured correlation, i.e. the probability that there is no timing noise present in the sample, was then computed with a permutation test (using 10,000 shuffles of the timing residuals). The probable presence of non-white noise in the sample is considered significant if the probability of chance occurrence is lower than 1\% (equivalent to a 99\% confidence level). The results are discussed in Section 3 and probabilities are listed in Table 1.  While this method is useful as a quick test to determine whether correlations are present, it is more thorough to compute the autocovariance at all lags, as presented in the next section.

\subsection{Weighted autocovariance at all lags}

For the following two methods (see Sections 2.2.1 and 2.2.2), we compute the weighted autocovariance of the daily-averaged timing residuals at all lags using 50-day time bins, separately for each pulsar and each large frequency band of the NANOGrav dataset \citep{demorest}. The weighted autocovariance of the daily-averaged timing residuals $R_i$, measured at times $t_i$ with the corresponding errors $\delta_i$, is computed as follows:
\begin{equation}
cov(\tau) = \sideset{}{_{i,j}}\sum_{|t_j\!-\!t_i\!-\!\tau|<\frac{\Delta\tau}{2}}\!\! \frac{w_iw_j}{W(\tau)} R_i \, R_j,
\end{equation}
where the sum runs over all pairs of residuals in a $\Delta\tau=50$\,day span around the time lag $\tau$, the individual
weights are $w_i=1/\delta_i^2$, and the sum of weights in a bin is
\begin{equation}
W(\tau) = \sideset{}{_{i,j}}\sum_{|t_j\!-\!t_i\!-\!\tau|<\frac{\Delta\tau}{2}}\!\! w_iw_j.
\end{equation}
The error $\sigma(\tau)$ on the autocovariance $cov(\tau)$ is computed as follows:
\begin{equation}
\sigma(\tau) = \sqrt{\frac{1}{W(\tau)}}.
\end{equation}

\subsubsection{Exponential Fit}

The power spectrum $P(f)$ of timing noise can be modeled as the sum of red noise and white noise components as follows:
\begin{equation}
P(f) = A f^{-\alpha}+B,
\end{equation}
where $f$ is the frequency of timing residuals, red noise is modeled by a power law with amplitude A and index $-\alpha$, and $B$ is the white noise component.
In \citet{coles}, however, the red noise is modeled using a similar power law but with a corner frequency $f_c$ (taking into account the fact that very low frequencies tend to be flat before decreasing as a power law):
\begin{equation}
P(f)=\frac{A}{\left(1+\left( f/f_c \right)^2 \right)^{\alpha/2}} + B
\end{equation}
The autocorrelation function of a time series (in this case the timing residuals) is the Fourier transform of the power spectrum. In this case the autocorrelation function is a K-Bessel function of the autocorrelation lag $d$:
\begin{equation}
C(d) = \frac{A}{\Gamma(\alpha/2)} 2^{1-\alpha/2} \left( \frac{1}{f_c^2} \right)^{-(\alpha+1)/4} \, |d|^{(\alpha-1)/2} \, K_{(1-\alpha)/2} \left( \frac{|d|}{\sqrt{1/f_c^2}} \right) + \sqrt{2 \pi} \, B \, \delta(d)
\end{equation}
In the case when $\alpha=2$ however, this expression simplifies dramatically, and takes the form of a decaying exponential:
\begin{equation}
C(d)=C_0 \, e^{-d/\tau} + \sqrt{2 \pi} \, B \, \delta(d),
\end{equation}
where $C_0=\sqrt{\pi/2} \, A f_c$ and $\tau=\sqrt{1/f_c^2}$, so one can easily compute $P(f)$ from $C(d)$.
This is the function described in Coles et al. (2011) and used in the TEMPO2 ``SpectralModel" plugin to fit for the autocorrelation function of timing residuals in pulsars with weak red noise. Coles et al. (2011) suggest that in the case of dominant red noise, it is not adequate to calculate the autocorrelation directly from timing residuals, and they have developed an algorithm using the Cholesky method to first estimate a power spectrum. However in the case of weak red noise, they calculate the autocorrelation function directly in the time domain, and fit a decaying exponential to the autocorrelation function, as seen in equation (8). We note that there is nothing physically special about the $\alpha=2$ case  (it is not in any way justified by neutron star physics), rather it provides an ``easy-to-calculate'' solution. 

Since (as seen from the overall shape of timing residuals), all NANOGrav pulsars in our data set exhibit at most weak red noise, we proceed in this way, fitting the decaying exponential of equation (8) to the autocovariance of timing residuals of equation (2). We note that the TEMPO2 plugin fit only minimizes the goodness-of-fit with respect to one parameter, the best lag $\tau$. We instead use a weighted non-linear least squares fit, minimizing with respect to both variables, the best lag $\tau$ and the best amplitude $C_0$, taking into account the errors in the autocovariance shown in equation (4). Best parameters were found using the ``optimize.curve-fit" function in Scientific Python, which makes use of the Levenburg-Marquadt algorithm. Parameter errors were found from the covariance matrix of estimated parameters. Results for the amplitude $C_0$ are listed in Table 1.

\subsubsection{Delta-function statistic}

For this method, we use the autocovariance at all lags computed in equation (2). In the case of white noise, we expect a very small value of the autocovariance at all non-zero lags. In fact, in the absence of non-white noise, we expect the autocovariance as a function of time lag to look like a delta-function (positive value at zero-lag and zero values for all non-zero lags). In order to estimate the presence of non-white noise, we calculate a goodness-of-fit $\chi^2$ showing the deviation from zero of the autocovariance points $cov(\tau)$ from equation (2) with the variance $\sigma(\tau)^2$ shown in equation (4):
\begin{equation}
\chi^2 = \frac{1}{T-1} \sum_{\tau \neq 0}{\frac{cov(\tau)^2}{\sigma(\tau)^2}},
\end{equation}
where T corresponds to the timespan of observations in days (and equals the maximum time lag between observations). Note that this $\chi^2$ is actually a reduced $\chi^2$ and corresponds to the goodness of fit of the autocovariance of timing residuals, and is not related to the $\chi^2$ of the parameter fit in TEMPO2.

\subsubsection{Time factor}

Additionally, we are interested in the amplitude of the non-white noise as compared to the amplitude of white noise. This can help us determine how many more observations would be needed for the RMS of timing residuals to be significantly influenced by the presence of non-white noise. Since the RMS of white-noise dominated residual averages is expected to decrease as $1/\sqrt{N}$ where N is the number of timing observations \citep{handzo}, we expect the variance of white noise (more specifically the uncertainty in the mean TOA averaged over the entire data set) to behave as follows:
\begin{equation}
\sigma_w^2 = \frac{\sigma^2_0}{N},
\end{equation}
where $\sigma_0$ is the unweighted RMS of timing residuals. The boundary between white noise-dominated and red noise-dominated residuals is defined by $\sigma_w=\sigma_r$, where $\sigma_w$ is the level of white noise and $\sigma_r$ is the level of red noise (calculated here using the absolute value of the maximum value of the autocovariance at non-zero lags). We can express this boundary in terms of $\sigma_0$ and $N$ instead: $\sigma_w^2=\sigma^2_0/N = \sigma_r^2$. We use this to define a ``time factor":
\begin{equation}
\text{Time Factor}=N=\frac{\sigma^2_0}{\sigma^2_r}
\end{equation}
This time factor gives us an estimate of the maximum factor by which we can multiply our current number of observations without residuals becoming dominated by red noise. In other words, we can still achieve significant gain in a PTA by increasing the number of observations up to N times what is currently being done. This is possible if we increase the number of observations within the same overall time span (here T=5 years), i.e. if we increase the cadence by a factor of N (since we expect red noise to become significant at longer time spans, extending the time span of observations might instead increase the amount of timing noise). 
The goodness-of-fit $\chi^2$, computed from equation (9), and time factors, computed from equation (11), were obtained for the NANOGrav dataset and listed in Table 1. 

\subsubsection{Possible issues}

It is important to note that we are using data which have been irregularly sampled.  \citet{chatfield} notes that the calculations of autocorrelation and autocovariance of non-uniformly sampled data are not mathematically rigorous. Additionally, there are issues with non-stationarity: after parameter fitting in ~TEMPO2, the timing data may no longer be stationary. Therefore expecting a power spectrum of timing residuals of the form shown in equations (5) or (6) may not be adequate. Keeping in mind these reservations, we calculate the autocovariance of timing residuals in order to estimate the presence of red noise in our data sample.

\section{Results}

Using the procedures outlined in Section 2, we analyzed the 17 pulsars of \cite{demorest}. We present in Table 1 the results of each method for each pulsar at each frequency band.

\textit{Weighted RMS level.}
The amplitude of the overall noise is characterized by the RMS of timing residuals. The lowest RMS (0.011 $\mu s$) is achieved by PSR J1909-3744 at 800 MHz, while the RMS of PSR B1953+29 at 1400 MHz is as high as 1.863 $\mu s$. As mentioned in Section 1, a PTA needs about 20 pulsars with an RMS of 0.1 to 0.5 $\mu s$ in order to detect GWs (Jenet et al, 2005). 
Since the RMS of timing residual averages is expected to decrease as $1/\sqrt{N}$ (where N is the number of observations), observing pulsars for longer times or increasing the cadence will lower the RMS of all pulsars, provided the amplitude of white noise is dominating the amplitude of red noise that may or may not be present. We now investigate the amplitude of red noise in NANOGrav pulsars using the statistics we described in Section 2.

\textit{Nearest-neighbor correlations.}
Nearest-neighbor correlation results are presented in Table 1 in the ``prob(\%)'' column. The presented number is the probability that the data sample is consistent with white noise. Small probabilities violate the null hypothesis, i.e. suggest that the data sample is \textit{not} consistent with white noise. For example, we would find significant evidence of non-white noise for prob $<1 \%$. Using this statistic, we find that none of the NANOGrav pulsars exhibit significant nearest-neighbor correlations. It is difficult to draw conclusions from this statistic alone, and we now focus our attention on methods that look at correlations at all lags.

\textit{Exponential Fit method.}
Using the method outlined in Section 2.2.1, we computed autocovariance values vs. lag and fit the decaying exponential (equation 8) to these values, giving us best-fit amplitudes $C_0$. In Table 1 we show the best-fit $C_0$ values and their uncertainties. We find that the fits for most pulsars are consistent with a zero amplitude, i.e. $C_0=0$, meaning the fit to an exponential is not a good fit and no correlation is observed according to this statistic, indicating the absence of red noise in the data sample. Only PSR J1918-0642 at 1400 MHz shows a deviation of more than one standard deviation, but this is not considered to be significant.

We also show plots of autocovariance vs. lag for several pulsars: J1012+5307 (which is clearly white) in Figure 1, J1643-1224 in Figure 2, and J1910+1256 in Figure 3. The latter two both exhibit weak red noise. The red box at zero-lag corresponds to the unweighted autocovariance at zero-lag, which is a measure of the white noise. In green we have plotted the exponential fit. We observe that while pulsars such as J1643-1224 and J1910+1256 exhibit significant structure in their plots of autocovariance vs. lag, these structures do not follow the expected exponential decay as prescribed in equation (8), i.e. the decaying exponential does not appropriately model the autocovariance data observed in these pulsars. More generally, we find that for all NANOGrav pulsars, the structure of the correlation function is not well-fitted by a decaying exponential. This is not unexpected, since the $\alpha=2$ spectral index is just one special case, and ideally one would want to fit for various spectral indices. One therefore needs to keep in mind that, when using the ``SpectralModel'' plugin in the case of weak red noise, the plugin uses this exponential function even though it may not be an appropriate fit for the autocovariance.

\textit{Delta-function statistic method.}
In the case of white noise, we expect the autocovariance as a function of lag to look like a delta function. As described in Section 2.2.2, we computed a goodness-of-fit $\chi^2$ (shown in Table 1) of the autocovariance values as compared to a delta function. We consider small values of $\chi^2$, specifically $\chi^2 < 9$ to be consistent with white noise. Values above 9 suggest that there is some non-zero structure in the plot of autocovariance vs. lag. We highlight in yellow $\chi^2$ values between 9 and 50, and in red values of $\chi^2 > 50$. We find that PSRs J1640+2224, J1643-1224, J1744-1134, J1853+1308, J1909-3744, J J1910+1256, J1918-0642 and J2317+1439 all exhibit a high $\chi^2$, therefore showing significant non-zero structure in the plot of autocovariance vs. lag. 

The $\chi^2$ measures how the autocovariance at non-zero lags deviates from zero, however it does not take into account the autocovariance value at zero-lag. In order to get a sense of whether any of these deviations is significant, we need to look at the ratio of the amplitude of white noise (the zero-lag value) vs. the amplitude of non-white noise. As discussed in Section 2.2.3, this ratio is also a measure of how much we can increase the cadence of observations without red noise becoming dominant, it is therefore called the ``time factor'' (equation 11). Results are shown in Table 1. A high time factor is consistent with white noise, while a low time factor suggests that the amplitude of the non-white noise is significant. For small $\chi^2$ (we choose $\chi^2<3$) , the data sample is consistent with white noise, and the value of the time factor is not applicable. In these cases we write ``NA'' for the time factor values.

By combining the results of the $\chi^2$ and time factor tests, which complement each other, we can better identify those pulsars which exhibit non-white noise (when both tests point to non-white noise: a high $\chi^2$ and a low time factor). The results are indicated by the coloring (red and yellow) in the table. We highlight in red cases for which the $\chi^2$ value is very high and the time factor is smaller than 2, while we highlight in yellow cases where the $\chi^2$ value is very high and the time factor is between 2 and 6. We find that only four of the pulsars with a high $\chi^2$: PSRs J1643-1224, J1910+1256, J1918-0642 and J2317+1439 also have a low time factor. Instead a pulsar like PSR J1744-1134, which has a very high $\chi^2$ of 849, also has a high time factor (179), making its non-zero structure irrelevant when compared to the amplitude of the white noise. We also note that unlike J1643-1224 and J1910+1256, which exhibit irregular patterns in the timing residuals \citep{demorest}, the timing residuals of J1918-0642 are much more regular. We will therefore place J1918-0642 in the ``borderline red" category rather than the definitely red one. 

To summarize this method: in the case of white noise, we expect a low $\chi^2$ (meaning the autocovariance data are consistent with a delta-function) and a high time factor (meaning that any amount of non-white noise is suppressed compared to the level of white noise). In the case of significant red noise,  we expect a high $\chi^2$ and a low time factor. This is the case for J1643-1224 ($\chi^2$ of 762 and time factor of 1.16). Additionally, we have pulsars exhibiting both a high $\chi^2$ and a high time factor. In this case we conclude that there is structure in the autocorrelation function, but that it is insignificant when compared to the level of white noise. 

\begin{figure}[HTP]
\begin{center} 
\includegraphics[width=5in]{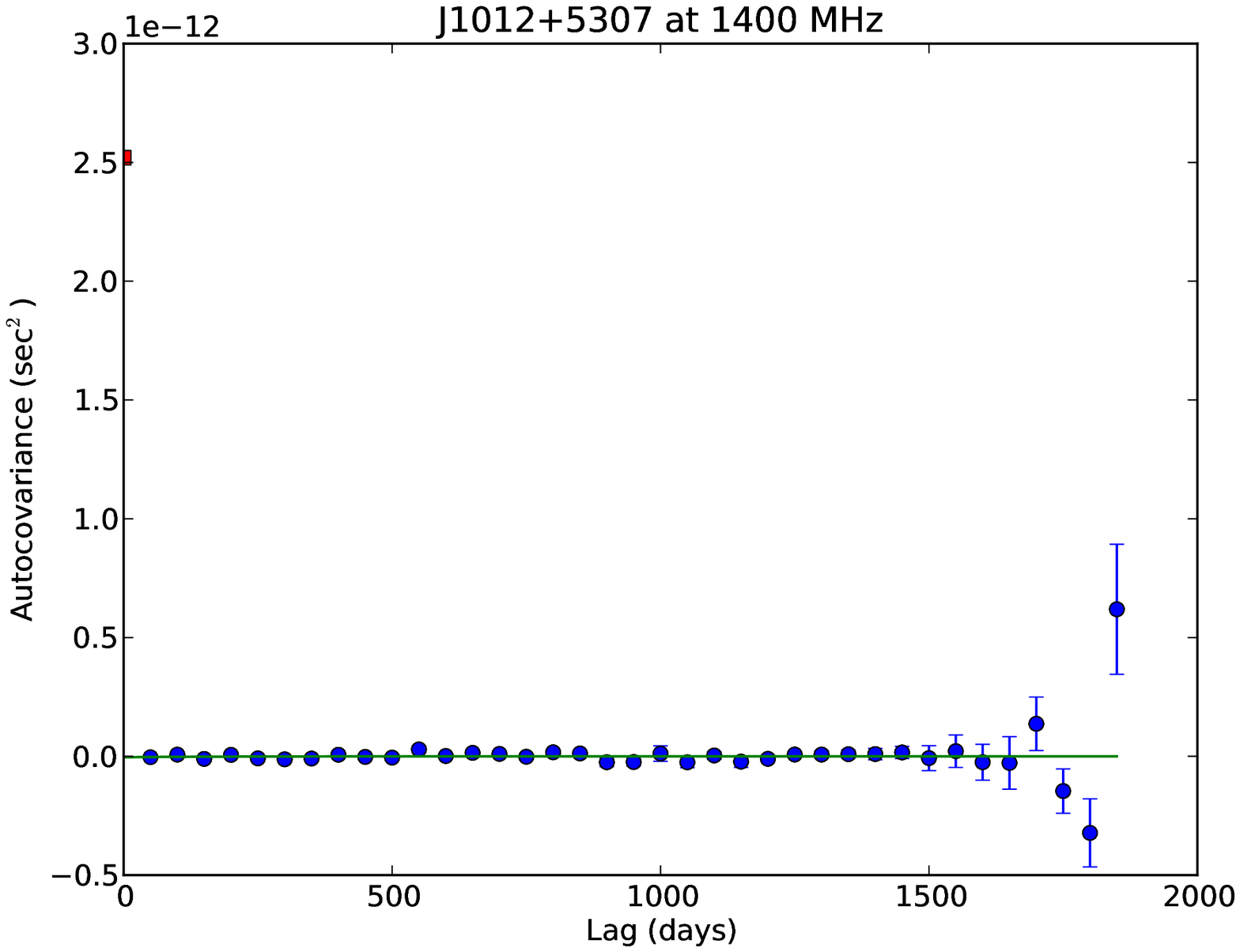} 
\end{center}
\caption{Autocovariance vs. lag for PSR J1012+5307: a ``white'' pulsar. The red box at zero-lag is the unweighted autocovariance at zero-lag and is a measure of the white noise. The green line corresponds to the fitted decaying exponential of equation (8).}
\end{figure}
\begin{figure}[HTP]
\begin{center} 
\includegraphics[width=5in]{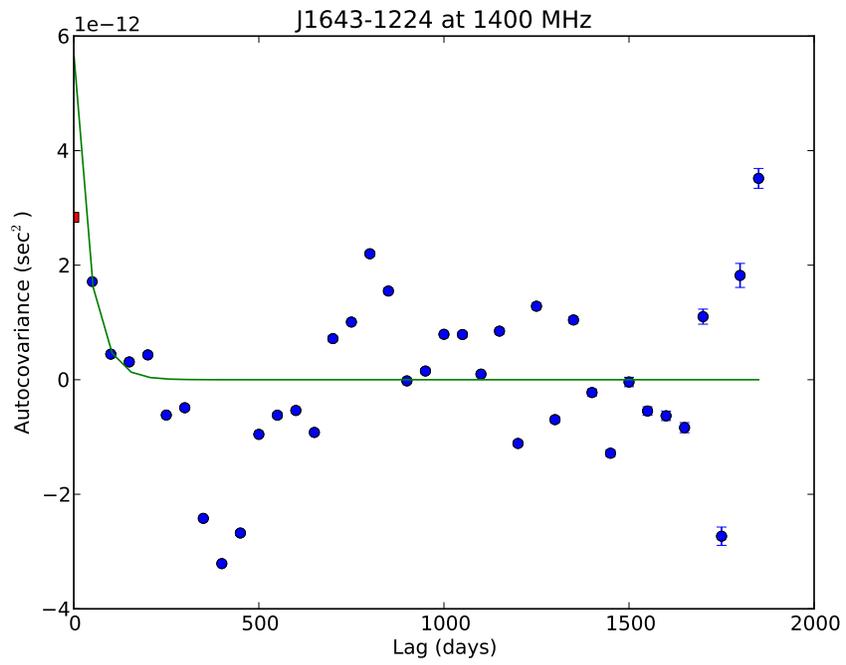} 
\end{center}
\caption{Autocovariance vs. lag for PSR J1643-1224: a ``red'' pulsar showing structure and a small time factor.}
\end{figure}
\begin{figure}[HTP]
\begin{center} 
\includegraphics[width=5in]{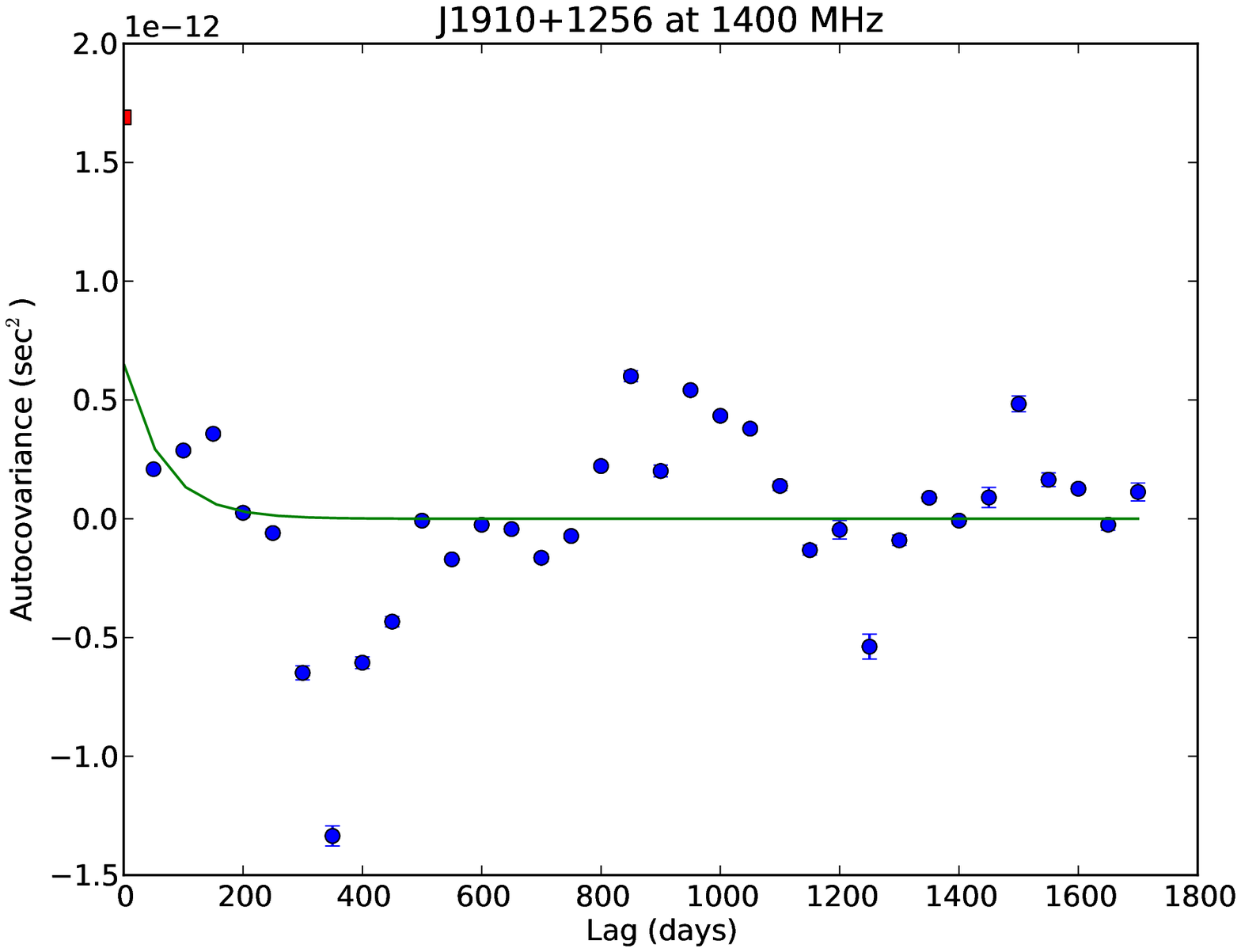} 
\end{center}
\caption{Autocovariance vs. lag for PSR J1910+1256: a ``red'' pulsar showing structure and a small time factor.}
\end{figure}

\textit{Advantages and limitations of each method}

In this paragraph, we explore the advantages and limitations of each method, and how they complement each other to create a coherent picture of the noise in MSPs. As mentioned earlier, the nearest-neighbor correlation is a good first estimator for determining the presence of non-white noise in pulsar timing residuals. If the data show significant correlation, this should appear at adjacent lags. However in the case of NANOGrav pulsars, the lack of correlation at adjacent lags indicates that either the NANOGrav pulsars are overwhelmingly white, or that this method is not adequate. It needs to be supplemented by a more quantitative approach to determine the amount of observed correlated noise.  The delta-function statistic provides a quantitative way to determine how consistent the autocovariance is with a delta function, i.e. how consistent the data are with white noise. This permits us to detect any non-white ``structure" in the plot of autocovariance vs. time lag. Additionally, the ``time factor'' tells us how significant or insignificant the amplitude of non-white noise is, as compared to the amplitude of white noise. Fitting a decaying exponential to the autocovariance is similar and complementary to the delta-function statistic. A positive amplitude of the exponential would show a certain amount of ``structure'' in the autocovariance data. However we observe in the plots of autocovariance vs. lag that this particular fit does not model the observed structures in the autocovariance of NANOGrav pulsars. Looking at each of these statistics, we can make conclusions about the presence of noise in NANOGrav pulsars.

\section{Summary and Implications}

We have characterized the timing noise in 17 NANOGrav pulsars using 3 different methods: one involving the computation of the nearest-neighbor correlation of the timing residuals; a second one involving the fitting of a decaying exponential to the autocovariance of the timing residuals as a function of time lag, and a third method involaving the fitting of a delta function to the same autocovariance data while also comparing the level of non-white noise to the level of white noise. Gathering results from all 3 methods, we conclude that all 17 pulsars exhibit dominant white noise in our observations. Seven pulsars only exhibit white noise:  PSRs J0030+0451, J0613-0200, J1012+5307, J1455-3330, J1600-3053, J1713+0747, and B1855+09. PSRs J1640+2224, J1744-1134, J1853+1308, J1909-3744, J1918-0642, J2145-0750, J2317+1439 and B1953+29 are considered borderline, as they show some evidence for correlations, but the results are not consistent across all statistics. Finally, PSRs J1643-1224 and J1910+1256 clearly show evidence for weak red noise. We note that \citet{demorest} and J. Ellis et al. (2013, in preparation) also found evidence of red noise in J1643-1224 and J1910+1256. We also note that J1910+1256 was observed at only one frequency band, and so DM variations were not fitted for. Also, J1640+2224 has been known to show strange behaviors, possibly because its orbital period is close to half a year, therefore orbital sampling and fitting may not be optimal. 
\begin{table}
\begin{center}
\begin{tabular}{ l c  c  c  c  c  c  r  }
\hline
Pulsar & Freq & Obs & $\omega$RMS ($\mu s$) & prob($\%$) & $C_0(10^{-13} s^2)$ & $\chi^2$ &  Time Factor  \\
\hline
J0030+0451 & 400 & AO & 0.019 & 72.3 & $\text{no fit}$ & 0.06  & NA   \\
J0030+0451 & 1400 & AO & 0.327 & 5.5 & $0.2 \pm 0.4$ & 0.96 &  NA   \\
J0613-0200 & 800 & GBT & 0.021 & 59.3 & $\text{no fit}$ & 0.001 & NA    \\
J0613-0200 & 1400 & GBT & 0.520 & 46.5 & $\text{no fit}$ & 5.8 & \yellow{$5 \pm 2$}   \\
J1012+5307 & 800 & GBT &  0.191 & 75.7 & $0.01 \pm 0.2$ & 0.005 & NA  \\
J1012+5307 & 1400 & GBT &  0.344 & 37.4 & $-0.04 \pm 2$ & 1.3 & NA   \\
J1455-3330 & 800 & GBT &  0.344 & 58.6 & $\text{no fit}$ & 0.02 & NA   \\
J1455-3330 & 1400 & GBT &  1.077 & 34.9 & $\text{no fit}$ &  0.34 & NA  \\
J1600-3053 & 800 & GBT & 0.208 & 32.7 & $\text{no fit}$ &  0.05 & NA  \\
J1600-3053 & 1400 & GBT & 0.135 & 86.5 & $\text{no fit}$ & 0.07 & NA   \\
J1640+2224 & 400 & AO &  0.057 & 23.2 & $\text{no fit} $ & 0.08 & NA   \\
J1640+2224 & 1400 & AO & 0.601 & 30.6 & $3 \pm 35$ & \red{385} & $79 \pm 4$   \\
J1643-1224 & 800 & GBT & 0.585 & 17.3 & $2 \pm 52$ & \yellow{9.5} & \yellow{$3 \pm 2$}    \\
J1643-1224 & 1400 & GBT & 1.880 & 31.7 & $58 \pm 160$ & \red{762} & \red{$1.16 \pm 0.06$}   \\
J1713+0747 & 800 & GBT & 0.091 & 21.3 & $0.02 \pm 0.03$ & 1.0 & NA   \\
J1713+0747 & 1400 & GBT/AO & 0.025 & 24.3 & $\text{no fit}$ & \yellow{15} & $56 \pm 4$   \\
J1713+0747 & 2300 & AO & 0.039 & 40.6 & $\text{no fit}$ & 6.7 &  $395 \pm 154$   \\
J1744-1134 & 800 & GBT & 0.140 & 25.0 & $\text{no fit}$ & 4.0 & $107 \pm 65$   \\
J1744-1134 & 1400 & GBT & 0.229 & 91.8 & $0.1 \pm 6$ & \red{849} & $179 \pm 40$   \\
J1853+1308 & 1400 & AO & 0.270 & 6.5 & $\text{no fit}$ & \red{220} & $15.8 \pm 0.9$   \\
B1855+09 & 400 & AO & 0.279 & 26.5 & $\text{no fit}$ & 0.26 & NA  \\
B1855+09 & 1400 & AO & 0.101 & 54.4 & $0.2 \pm 2$ & \yellow{16.9} & $21 \pm 4$   \\
J1909-3744 & 800 & GBT & 0.011 & 69.4 & $\text{no fit}$ & 2.2 & NA   \\
J1909-3744 & 1400 & GBT & 0.048 & 78.5 & $0.2 \pm 7$ & \red{684} & $14.9 \pm  0.4$   \\
J1910+1256 & 1400 & AO & 0.709 & 20.4 & $7 \pm 17$ & \red{294} & \yellow{$5.3 \pm 0.3$}     \\
J1918-0642 & 800 & GBT & 0.129 &  92.3 &  \text{no fit} & 1.9  & NA   \\
J1918-0642 & 1400 & GBT & 0.211 & 28.1 & $3.2 \pm 1.4$ & \red{950} & \yellow{$4.88 \pm 0.08$}  \\
B1953+29 & 1400 & AO & 1.863 & 44.3 & $\text{no fit}$ & \yellow{15.5} & \yellow{$3.0 \pm 0.7$}    \\
J2145-0750 & 800 & GBT & 0.068 & 63.9 & \text{no fit} & \yellow{10} & \red{$0.33 \pm 0.09$}    \\
J2145-0750 & 1400 & GBT & 0.494 & 1.8  & $3.2 \pm 9$ & \yellow{12.4} &  $20 \pm 4$ \\
J2317+1439 & 300 & AO & 0.369 & 95.2 & $-0.7 \pm 6$ & \red{57} & \yellow{$3.3 \pm 0.4$}    \\
J2317+1439  & 400 & AO & 0.153 & 82.9 & \text{no fit} & 5.5 & \yellow{$3.3 \pm 0.7$}    \\
\hline
\end{tabular}
\end{center}
\caption{Noise analysis for each NANOGrav pulsar at each frequency band (in MHz) observed at the Arecibo (AO) and Green Bank (GBT) telescopes over a 5-year period. We show the weighted RMS (representing the ``white'' noise component); results of the nearest-neighbor correlation (``prob($\%$)": probability that the residuals are consistent with white noise); the amplitude $C_0$ of an exponential fit to the autocovariance; a goodness-of-fit ($\chi^2$) of the autocovariance values compared to a delta-function;  and the ``factor'' needed to reach the level of red noise (indicating how many more observations we can make and still expect the RMS to decrease as $1/\sqrt{N}$.) Yellow colors indicate a slight deviation, and red a significant deviation from non-white noise. See Section 3 for a detailed analysis.}
\end{table}

One important implication to note is that while we have evidence for red noise in these 2 pulsars, they exhibit at most ``weak'' red noise rather than dominant red noise. This means that white noise is still dominant; we can increase the cadence of our observations and still expect the RMS to decrease as $1/\sqrt{N}$. Increasing the cadence of observations is a key strategy to reduce the RMS of timing residuals, making our PTA a more sensitive instrument for GW detection. Perhaps the most important result of this paper is that regardless of the category into which we have placed each pulsar, the cadence may still be increased before red noise starts to dominate. Therefore it is to our advantage to increase the observing time on every NANOGrav pulsar. This idea is further developed in the work of Handzo et al (2013).

Finally, we comment on the advantages and limitations of each of these 3 methods. The  computation of the ``nearest-neighbor" correlation is useful as a first estimate for the presence of correlations in timing residuals, however, it does not give a complete description of all the possible correlations present, and it is difficult to determine time scales associated with the noise process. The delta-function statistic uses the entire autocovariance function to determine the whiteness of the data. Through the computation of the ``time factor'', it also provides a way to determine the amplitude of the non-white noise component as compared to the white noise component. The fitting of the decaying exponential to autocovariance values, motivated by its equivalent power law form in the frequency domain, provides a specific, quantifiable way to characterize red noise. However it does not appear to appropriately fit autocovariance data for NANOGrav pulsars.

A lot of work still needs to be done to understand the nature of timing noise in pulsars, which is crucial to detecting GWs with PTAs. Indeed, if we don't understand the noise in pulsars, we cannot make good choices regarding the directions of PTAs, such as how long to integrate, which pulsars to observe more often, etc. We need to determine the cause of red noise (spin noise or interstellar medium); properly characterize the red noise either using frequentist or bayesian methods; understand why different PTAs observed different noise characteristics for the same pulsars. All of these goals will be aided with an understanding of how DM variations are performed in different PTAs (a current IPTA project). Other methods are currently being investigated to characterize timing noise in pulsars, including a zero-crossing analysis by Y. Wang et al (2013, in preparation), bayesian methods by J. Ellis et al. (2013, in preparation), and will be synthesized in the work of J. Cordes et al. (2013, in preparation). All timing residuals for the NANOGrav dataset can be found in \citet{demorest}.

\section{Acknowledgements}
We would like to thank Bill Coles, George Hobbs, Daniel Yardley, Jim Cordes, Justin Ellis, Xavi Siemens, Brian Christy, Joe Romano and Ben Stappers for very useful discussions. This project has been supported by NSF AST CAREER 07-48580 and NSF PIRE 0968296. Pulsar research at UBC is supported by an NSERC Discovery Grant. 

\bibliography{timingnoise}

\end{document}